\documentclass[
    aps,
    pra,
    reprint,
    amsmath,amssymb,
    superscriptaddress,
    floatfix,
    longbibliography
]{revtex4-2}

\usepackage{graphicx}
\usepackage{bm}
\usepackage{siunitx}
\usepackage{xcolor}
\usepackage{placeins}

\graphicspath{{figures/main/}{figures/supplement/}{Figures/}{SI_Figures/}}

\newcommand{\nlp}{\bar{n}_{\mathrm{LP}}}
\newcommand{\aref}{\alpha_{\mathrm{ref}}}
\newcommand{\nhat}{\hat{n}}
\newcommand{\Tone}{T_{1}}
\newcommand{\Tphi}{T_{\phi}}
\newcommand{\pk}{p_{k}}
\newcommand{\ket}[1]{\lvert #1 \rangle}
\newcommand{\bra}[1]{\langle #1 \rvert}
\newcommand{\melem}[3]{\langle #1 \rvert #2 \lvert #3 \rangle}
\newcommand{\ncrit}{n_{\mathrm{crit}}}
\newcommand{\nres}{n_{\mathrm{res}}}

\begin{document}

\title{Persistent level-resolved transmon population after dispersive readout, and the limits of its open-system description}

\author{Sinan Bu\u{g}u}
\email{sbugu@ncsu.edu}
\affiliation{Department of Physics and Astronomy, North Carolina State University, Raleigh, North Carolina 27695, USA}
\affiliation{BuQuLab LLC, Winston-Salem, North Carolina 27101, USA}

\date{\today}

\begin{abstract}
We calculate the level-resolved transmon population remaining after dispersive readout with a converged open-system model. The drive and qubit--cavity coupling are treated in the excitation-conserving (rotating-wave) approximation, while the cavity is retained as a quantized mode in a frame displaced by its classical reference trajectory. At a low-power-equivalent readout occupation of $\nlp = 80$, the model leaves a persistent distribution across the first several transmon levels, with $p_{2}=1.02\times10^{-3}$ and $p_{3}=4.10\times10^{-5}$ at the end of readout. These populations also survive an additional drive-off wait. Over $73$~ns, each level decays by less than $3\%$ and follows its independently calculated relaxation rate rather than the near-factor-of-two decrease in the residual cavity photon number. Within the model, this behavior identifies the remaining population as bare-level occupation rather than coherent dressing. Channel-removal and bath-model controls attribute $p_{2}$ mainly to dephasing and $p_{3}$ and higher levels mainly to relaxation. This assignment is unchanged across the bath models tested, although the absolute populations vary by as much as about $15\%$. We also test the coherent drive truncation directly. In dissipation-free transmon-only benchmarks driven by the prescribed classical cavity trajectory, the nearest-neighbour RWA drive underestimates the full charge-operator result while the drive is on. At drive switch-off for $\nlp=80$, where the coherent populations are hundreds of times larger than the reported residual values, $p_{2}^{\mathrm{RWA}}/p_{2}^{\mathrm{full}}=0.905$ and $p_{3}^{\mathrm{RWA}}/p_{3}^{\mathrm{full}}=0.759$, with larger discrepancies at deeper levels. One structural source is a parity-allowed non-nearest-neighbour charge matrix element omitted by the nearest-neighbour drive operator. A naive beyond-RWA open-system extension using bare-basis Lindblad dissipators fails a zero-drive ground-state consistency test and is therefore used only as a diagnostic. We then implement a dressed-basis (secular) construction and confirm that, when initialized in the dressed ground state, it leaves that state stationary at zero drive. The result is a converged RWA-model prediction of readout-induced population together with direct tests showing where important parts of the approximation lose quantitative reliability.
\end{abstract}

\maketitle

\section{Introduction}\label{sec:intro}

Dispersive readout is the standard measurement method for most superconducting qubits. The qubit shifts the frequency of a coupled resonator in a state-dependent way, and the shift is inferred by driving the resonator and homodyning its output \cite{Blais2004,Blais2021,Krantz2019}. Readout has become both faster and more accurate over the past decade \cite{Walter2017,Jeffrey2014,Heinsoo2018}, in large part because quantum error correction requires repeated rounds of parallel measurement \cite{fowler2012surface,terhal2015quantum,krinner2022realizing,Acharya2025}. Short integration times and high assignment fidelity generally require more measurement photons, which pushes the readout toward stronger drive.

That stronger drive also moves the system away from the regime in which the usual dispersive description is most reliable. The dispersive model is a low-power expansion \cite{Gambetta2006,Clerk2010,boissonneault2009dispersive}. The resonator field ac-Stark shifts and dephases the qubit \cite{schuster2005ac}, and the approximation begins to degrade as the intracavity occupation approaches the critical photon number \cite{Gambetta2008}. Beyond this regime, population can leave the computational subspace. Measurement-induced state transitions populate higher transmon levels \cite{reed2010high,Sank2016_MIST,nesterov2024measurement}, and sufficiently strong drive can ionize the transmon into extended states \cite{Dumas2024,Shillito2022_Ionization,Verney2019_StructuralInstability,lescanne2019escape}. The dependence of these transitions on drive power, detuning, and offset charge is still not fully understood \cite{fechant2025offset,Swiadek2024}, and they remain an important limitation for fast readout \cite{Sunada2024,Spring2025}.

Most previous work has focused on what happens inside or immediately adjacent to the computational subspace, such as the assignment error caused by a transition or the drive power at which ionization begins. Less is known about the full higher-level population left by a complete readout pulse and whether that population remains after the pulse. This distinction matters for subsequent operations. While the drive is present, some higher-level population can be reversible dressing and return adiabatically when the drive is removed. Population that decoheres into genuine bare-level occupation is different: it persists, cannot be refocused by echo, and can seed errors in later operations \cite{Hazra2025,Bengtsson2024}. A single-shot fidelity does not show whether a readout leaves population in the second or third transmon level ($\ket{2}$ or $\ket{3}$) at the $10^{-3}$ or $10^{-5}$ level, nor whether that population survives \cite{danjou2021generalized}.

We calculate this level-resolved distribution directly. The readout mode is retained as a quantized field in a frame displaced by its classical trajectory, allowing higher-level populations to be resolved down to $10^{-6}$ while two representative readout powers are propagated through a complete pulse. The drive and qubit--cavity coupling are treated in the excitation-conserving (rotating-wave) approximation. This model preserves the joint vacuum exactly and can be converged numerically \cite{WisemanMilburn2010}, which lets us state clearly which conclusions belong to the model. Within it, readout leaves persistent population across the first several transmon levels. An additional drive-off wait distinguishes persistent bare-level occupation from coherent dressing: the populations follow the independently calculated level-relaxation rates rather than the cavity ring-down. Channel-removal and bath-model controls then identify the dissipative mechanism level by level and test whether that assignment depends on the bath model.

The numerical values are therefore reported explicitly as predictions of this model. At the drive strengths considered here, the excitation-conserving approximation is close to the edge of its usual domain: the strong-drive case is near the critical photon number, where beyond-rotating-wave effects associated with measurement-induced transitions become important \cite{Sank2016_MIST}. We test one part of the approximation directly by comparing the nearest-neighbour RWA drive with the full charge-operator drive in dissipation-free, transmon-only calculations, both at low occupation and at the $\nlp=80$ drive strength. This comparison reveals a specific structural omission, namely a parity-allowed non-nearest-neighbour charge matrix element absent from the nearest-neighbour drive operator. We do not assign the full discrepancy to this matrix element, nor do we claim that it is the only source.

A natural next step is to keep the counter-rotating terms and propagate the open-system dynamics with the full Hamiltonian, but the naive version of that extension is not physically consistent. Bare-basis Lindblad dissipators combined with counter-rotating coupling do not preserve the dressed ground state of the interacting Hamiltonian. Even with no drive, the model then populates the qubit because the true ground state is dressed while the bare-basis jump operators violate detailed balance \cite{Beaudoin2011}. This problem is well known when the coherent and dissipative parts of an open-system model are not constructed consistently \cite{WisemanMilburn2010,harrington2022engineered}. Agreement between two implementations of the same model does not detect it because both implementations share the same dissipator. We therefore use the beyond-rotating-wave open-system calculations only as diagnostics and require them to pass a zero-drive ground-state consistency test before any population can be interpreted physically. The appropriate repair is a dressed-basis dissipator construction \cite{Beaudoin2011,boutin2017resonator,Gautier2025}. Recent work on the crossover between Markovian and non-Markovian readout dynamics emphasizes the same broader point: the physical validity of an open-system readout model depends on the dissipator as well as the Hamiltonian \cite{Murch2024_NonMarkovian}.

The paper has two related aims. First, we give a converged but deliberately scope-limited prediction of readout-induced level population and its dissipative origin. Second, we identify where the approximation used for that calculation loses reliability and give a reproducible consistency test for the open-system extension. This complements our earlier treatment of readout as a stochastic channel optimized for certification throughput \cite{bugu2026beyond}, where the computational subspace was taken as given; here we focus on the population that leaves it. Section~\ref{sec:model} defines the model, displaced-frame propagation, and numerical scope. Section~\ref{sec:dynamics} presents the driven dynamics, final level distribution, and persistence test. Section~\ref{sec:mechanism} isolates the dissipative mechanism and checks its dependence on the bath model. Section~\ref{sec:limits} tests the coherent drive truncation in a transmon-only calculation. Section~\ref{sec:failure} then examines the failure of the naive beyond-rotating-wave open-system extension and the consistency test that exposes it.

\FloatBarrier
\section{Model and methods}\label{sec:model}

We treat a single transmon dispersively coupled to a readout cavity~\cite{Koch2007,Blais2021}, with the transmon written in the charge basis,
\begin{equation}
    H_{\mathrm{t}} = 4E_{C}(\nhat - n_{g})^{2} - E_{J}\cos\hat{\varphi},
    \label{eq:transmon}
\end{equation}
where $E_{C}/h = 0.25$~GHz and $E_{J}/h = 12.5$~GHz, giving $E_{J}/E_{C}=50$. We diagonalize this Hamiltonian on a truncated charge grid and retain its lowest $n_{\mathrm{lev}}$ eigenstates $\ket{k}$. The resulting bare $0$--$1$ transition frequency is $\omega_{01}/2\pi\simeq 4.735$~GHz. The readout mode has $\omega_{r}/2\pi = 6.5$~GHz and linewidth $\kappa/2\pi = 5$~MHz, corresponding to a transmon--cavity detuning of $\Delta/2\pi\simeq -1.765$~GHz. For charge coupling $g/2\pi = 100$~MHz, the critical photon number is $\ncrit = (\Delta/2g)^{2}\simeq 78$. The strong-drive case considered below is therefore close to $\ncrit$, not deep in the low-power regime, and its excitation-conserving results are treated as model predictions rather than exact values. We set the offset charge to $n_{g}=0$ for all production runs; at this value the charge-parity selection rules used in Sec.~\ref{sec:limits} are exact. The device parameters are summarized in Table~\ref{tab:params}.

\begin{table}[t]
\caption{Model parameters. Mode and coupling quantities are reported as frequencies ($\omega/2\pi$), while $E_{C}$ and $E_{J}$ are energies divided by $h$. The dispersive shifts $\chi_{g,e}$ are obtained from exact diagonalization of the full counter-rotating Hamiltonian and are used to define the displaced reference frame.}
\label{tab:params}
\begin{ruledtabular}
\begin{tabular}{lll}
Quantity & Symbol & Value \\
\hline
Charging energy      & $E_{C}/h$        & $0.25$~GHz \\
Josephson energy     & $E_{J}/h$        & $12.5$~GHz \\
Ratio                & $E_{J}/E_{C}$    & $50$ \\
Cavity frequency     & $\omega_{r}/2\pi$& $6.5$~GHz \\
Coupling             & $g/2\pi$         & $100$~MHz \\
Cavity linewidth     & $\kappa/2\pi$    & $5$~MHz \\
Relaxation time      & $\Tone$          & $100~\mu$s \\
Pure-dephasing time  & $\Tphi$          & $150~\mu$s \\
Ground-state pull    & $\chi_{g}/2\pi$  & $+4.758$~MHz \\
Excited-state pull   & $\chi_{e}/2\pi$  & $+2.648$~MHz \\
Critical photon no.  & $\ncrit$         & $\simeq 78$ \\
\end{tabular}
\end{ruledtabular}
\end{table}

\subsection{Displaced-frame open-system propagation}\label{sec:frame}

We propagate the dynamics in a frame displaced by the classical cavity amplitude $\aref(t)$, obtained from the driven, damped cavity equation of motion in the frame matched to the initial transmon state. In this frame only quantum fluctuations around the classical trajectory need to be represented, so the retained Fock cutoff $N_{f}$ can remain small even at high readout power. We denote the reduced transmon state by $\rho_{\mathrm{q}}(t)$ and the level populations by $\pk(t)=\melem{k}{\rho_{\mathrm{q}}(t)}{k}$.

The evolution obeys a Lindblad master equation,
\begin{equation}
    \dot{\rho}
    = -\tfrac{i}{\hbar}\,[H(t),\rho]
      + \kappa\,\mathcal{D}[a]\rho
      + \sum_{j<k}\Gamma_{k\to j}\,\mathcal{D}[\ket{j}\!\bra{k}]\rho
      + \mathcal{D}_{\phi}\rho ,
    \label{eq:lindblad}
\end{equation}
where $\mathcal{D}[L]\rho = L\rho L^{\dagger} -\tfrac12\{L^{\dagger}L,\rho\}$. The time-dependent Hamiltonian $H(t)$ contains the transmon of Eq.~\eqref{eq:transmon}, the retained cavity mode, the qubit--cavity coupling, and the programmed readout drive. The converged calculations use the excitation-conserving (RWA) model. In this model both the drive and the coupling are truncated to their excitation-number-preserving parts, and the charge-drive operator retains only nearest-neighbour raising transitions with $\Delta k = +1$. Section~\ref{sec:limits} tests the effect of this truncation.

Relaxation is distributed over all allowed downward transitions using charge-matrix-element weights, $\Gamma_{k\to j}\propto \lvert\melem{j}{\nhat}{k}\rvert^{2}$ for $j<k$, corresponding to a flat, or white, bath. The rates are normalized so that the $1\!\to\!0$ transition has rate $1/\Tone$. Pure dephasing $\mathcal{D}_{\phi}$ is represented by a linear diagonal operator normalized to give a $0$--$1$ coherence decay rate of $1/\Tphi$. In Sec.~\ref{sec:mechanism} we vary both constructions to determine whether the level-resolved conclusions depend on these bath choices.

\subsection{Readout protocol}\label{sec:protocol}

The programmed drive envelope $\varepsilon(t)$ has a $\sin^{2}$ ring-up for $0\le t<60$~ns, followed by a flat dwell and an abrupt switch-off at $t=260$~ns. We use two readout powers, parameterized by the low-power-equivalent occupation $\nlp$. The dwell amplitude is $\varepsilon_{0}/2\pi = 12.135$~MHz for $\nlp = 20$ and $24.270$~MHz for $\nlp = 80$, giving peak reference occupations of $19.94$ and $79.78$ photons. After switch-off, each simulation continues until the reference cavity occupation $\lvert\aref\rvert^{2}$ first falls below the absolute stop criterion of $10^{-3}$. This occurs at $t=575.2$~ns for $\nlp = 20$ and $t=619.3$~ns for $\nlp = 80$. Both simulations start in $\ket{1}$. Figure~\ref{fig:readout_dynamics} shows the drive, reference trajectory, and time-dependent level populations.

\subsection{Numerical control and scope discipline}\label{sec:scope}

We test convergence along four directions: the Fock cutoff $N_{f}$, integration timestep, number of transmon levels $n_{\mathrm{lev}}$, and charge-basis cutoff. Together these give seven refinement comparisons across the two readout powers (Fig.~\ref{fig:supp_convergence}). The largest change observed on any refinement axis is $\max_{k}\lvert\Delta \pk\rvert = 2.68\times10^{-5}$, which we use as a conservative global resolution threshold. For $p_{3}$, every refinement changes the population by at most $5.9\times10^{-7}$. Both $p_{2}$ and $p_{3}$ satisfy the convergence criteria. The remaining model uncertainty is dominated by the bath choice rather than by numerical convergence (Sec.~\ref{sec:mechanism}).

We separate the results into three categories, and each figure states which category applies. \emph{Converged RWA predictions} (Figs.~\ref{fig:readout_dynamics}--\ref{fig:channels} and \ref{fig:supp_convergence}) are obtained from the excitation-conserving open-system model, which preserves the vacuum exactly. The \emph{coherent comparisons} in Sec.~\ref{sec:limits} are dissipation-free, transmon-only calculations that test the nearest-neighbour drive truncation. Figure~\ref{fig:coherent_rwa}(a) gives the low-occupation benchmark, and a separate calculation tests the same truncation at the $\nlp=80$ drive strength. These calculations neither validate beyond-RWA open-system dynamics nor provide a correction factor for the RWA populations. The \emph{model-validity diagnostics} in Figs.~\ref{fig:vacuum_gate}(b,c) and \ref{fig:supp_nonrwa_diagnostic} are beyond-RWA open-system calculations that fail a physical-consistency test; none of their populations is interpreted as physical. We do not describe any calculation in this work as exact, and we do not infer a correction factor for the RWA populations.

\FloatBarrier
\section{Readout dynamics and the final level distribution}
\label{sec:dynamics}

\begin{figure*}[t]
    \centering
    \includegraphics[width=\textwidth]{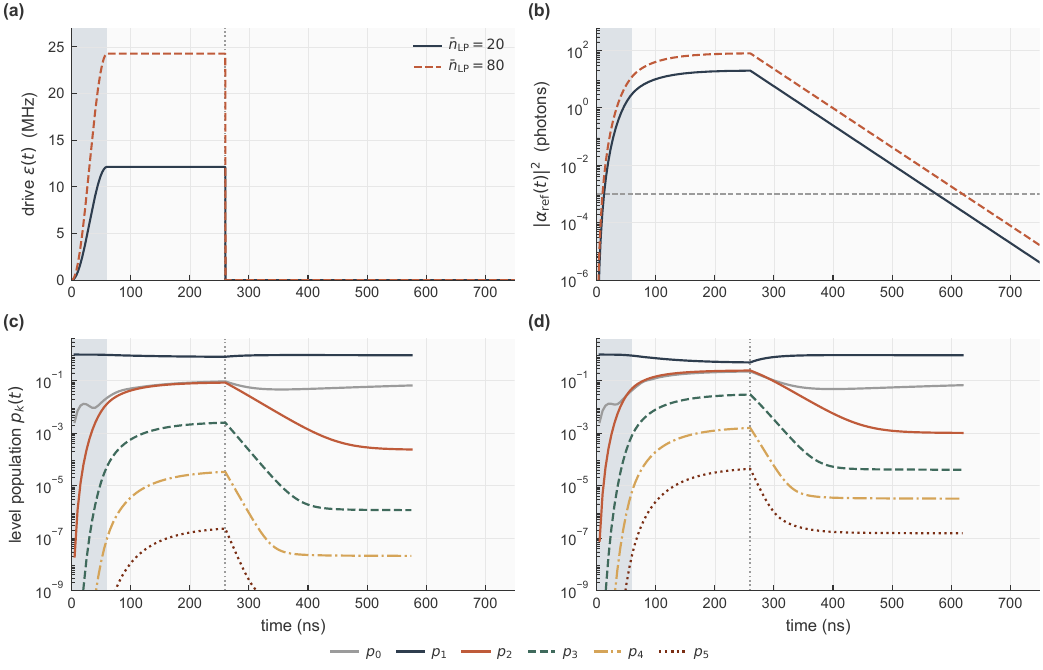}
    \caption{Readout dynamics at the two powers ($\bar{n}_{\mathrm{LP}}=20$, solid dark; $\bar{n}_{\mathrm{LP}}=80$, dashed light; both runs start in $|1\rangle$). (a)~Drive envelope $\varepsilon(t)$, with a $\sin^{2}$ ring-up over $0\le t<60$~ns (grey band), a flat dwell at $\varepsilon_0/2\pi = 12.135$~MHz for $\bar{n}_{\mathrm{LP}}=20$ or $24.270$~MHz for $\bar{n}_{\mathrm{LP}}=80$, and abrupt switch-off at $t=260$~ns (dotted line, also shown in panels (c) and (d)). (b)~Reference cavity occupation $|\alpha_{\mathrm{ref}}(t)|^{2}$ on a logarithmic axis. The dashed horizontal line marks the $|\alpha_{\mathrm{ref}}|^{2}=10^{-3}$ stopping criterion. (c),~(d)~Level populations $p_k(t)$ for $k=0$ to $5$ at $\bar{n}_{\mathrm{LP}}=20$~(c) and $\bar{n}_{\mathrm{LP}}=80$~(d), shown on a logarithmic scale from $10^{-9}$ to $4$. Levels above $k=5$ are below the displayed range and are included in Fig.~\ref{fig:final_populations}(a). All panels are converged results of the excitation-conserving model. The strong-drive result in~(d) should be interpreted together with the approximation test in Fig.~\ref{fig:coherent_rwa}.}
    \label{fig:readout_dynamics}
\end{figure*}

Figure~\ref{fig:readout_dynamics} shows the driven evolution. During the dwell, the reference cavity reaches occupations of tens of photons while population develops in transmon levels above the initial state. After switch-off, the reference amplitude rings down until it crosses the $10^{-3}$ stop criterion. Panels (c) and (d) show that the higher-level populations appear while the drive is on and do not return to zero when the drive is removed. We define the end of readout as the first post-switch-off crossing of $\lvert\aref\rvert^{2}=10^{-3}$ and evaluate the final populations at that time.

\begin{figure*}[t]
    \centering
    \includegraphics[width=\textwidth]{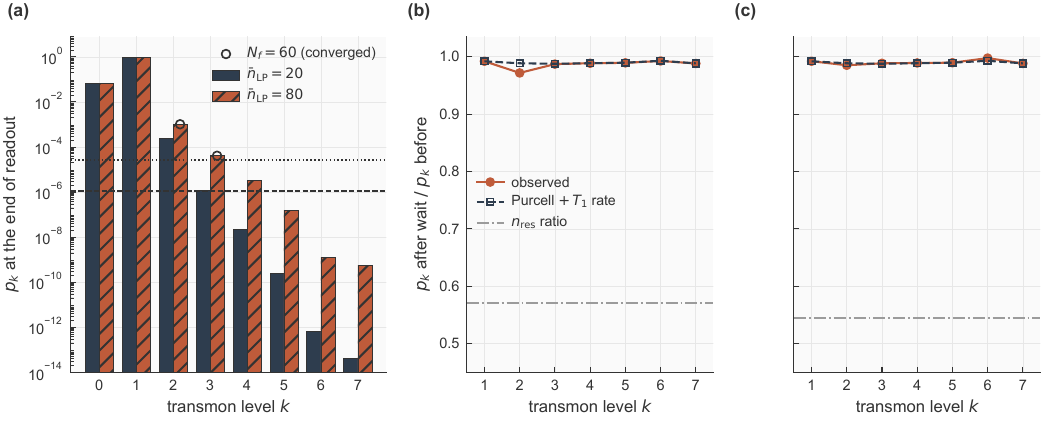}
    \caption{Final level distribution and persistence during an additional drive-off wait. (a)~Populations $p_k$ for $k=0$ to $7$ at the end of readout, defined by the first post-switch-off crossing of $|\alpha_{\mathrm{ref}}|^{2}=10^{-3}$, on a logarithmic axis. Solid dark bars correspond to $\bar{n}_{\mathrm{LP}}=20$ and hatched light bars to $\bar{n}_{\mathrm{LP}}=80$. Open circles show the converged $N_f=60$ values of $p_2$ and $p_3$ at $\bar{n}_{\mathrm{LP}}=80$. The dashed and dotted horizontal lines mark the level-specific and conservative global resolution thresholds; the global threshold is defined in Sec.~\ref{sec:scope} and the level-specific threshold in Sec.~\ref{sec:dynamics}. (b),~(c)~Ratio of each population after a further $73$~ns of drive-off evolution to its end-of-readout value at $\bar{n}_{\mathrm{LP}}=20$~(b) and $\bar{n}_{\mathrm{LP}}=80$~(c). Filled circles are the computed ratios, open squares are the ratios predicted from the independently calculated Purcell-plus-$T_1$ decay rate of each level, and the dash-dotted line is the residual-cavity-occupation ratio over the same interval.}
    \label{fig:final_populations}
\end{figure*}

The resulting distribution is shown in Fig.~\ref{fig:final_populations}(a). For $\nlp = 80$, the converged $N_{f}=60$ values are $p_{2}=1.0240\times10^{-3}$ and $p_{3}=4.1015\times10^{-5}$. The production-run values, $1.024\times10^{-3}$ and $4.036\times10^{-5}$, are visually indistinguishable from the converged values on the plotted scale but are not used as the converged headline values. At $\nlp = 20$, the corresponding populations are $2.388\times10^{-4}$ and $1.165\times10^{-6}$. We show two numerical floors rather than relying on a single convergence criterion. The level-specific threshold for $p_{3}$ is $1.139\times10^{-6}$, defined as the maximum of the per-level refinement residuals and the residual of an independent closed-system consistency check (two independent integrators of the dissipation-free dynamics), which sets the value $1.139\times10^{-6}$ used here. The $\nlp = 80$ value of $p_{3}$ is $36$ times this threshold. Relative to the more conservative global threshold of $2.680\times10^{-5}$, it is $1.5$ times the threshold. Therefore, both criteria place $p_{3}$ above the numerical floor at the higher power.

The additional drive-off wait tests whether these populations are bare-level occupation or transient coherent dressing. Panels (b) and (c) compare each level population after another $73$~ns of drive-off evolution with its value at the end of readout. If the population were tied to the residual cavity field, it would follow the dash-dotted line. Over the same interval, the residual occupation $\nres$, defined as the cavity photon number in the displaced reference frame and including a non-decaying dispersive floor, falls to $0.571$ of its pre-wait value for $\nlp=20$ and $0.545$ for $\nlp = 80$. This is a decrease of somewhat less than a factor of two. Bare-level population should instead decay according to the independently calculated Purcell-plus-$\Tone$ rate of each level, shown by the dashed line. The computed ratios follow this relaxation prediction, ranging from $0.972$ to $0.992$ at $\nlp = 20$ and from $0.985$ to $0.998$ at $\nlp = 80$. They are far from the cavity ratio; even the largest deviation from the per-level relaxation prediction, for $p_{2}$ at $\nlp = 20$, remains well separated from the cavity value. Within the model, the higher-level population is therefore persistent bare-level occupation rather than coherent dressing that disappears when the drive is removed.

\FloatBarrier
\section{Dissipative origin within the excitation-conserving model}
\label{sec:mechanism}

\begin{figure*}[t]
    \centering
    \includegraphics[width=\textwidth]{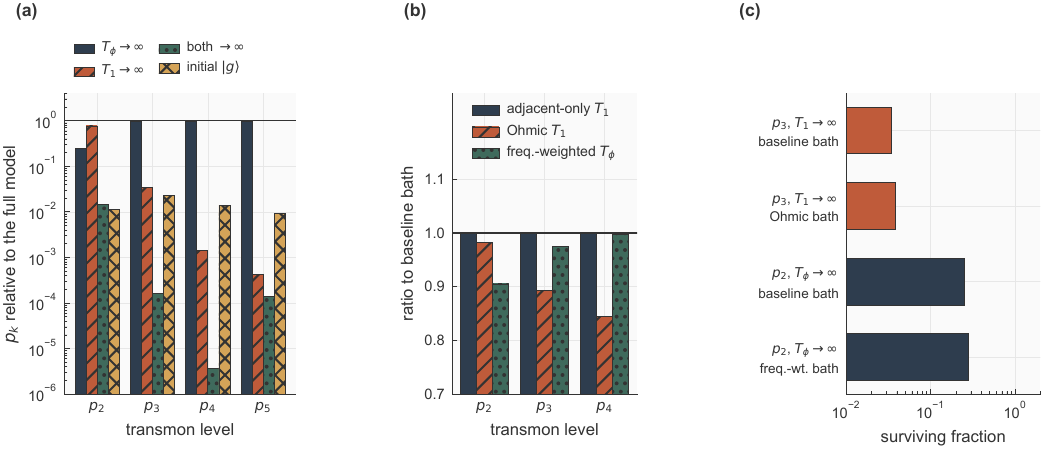}
    \caption{Channel-removal and bath-model tests at $\bar{n}_{\mathrm{LP}}=80$ and offset charge $n_g=0$. (a)~Populations for $k=2$ to $5$ after removal of one dissipative channel, normalized to the full-model value and shown on a logarithmic axis; the line at unity marks the full model. Pure dephasing ($T_\phi\to\infty$), relaxation ($T_1\to\infty$), or both are removed in separate propagations rather than by post-hoc subtraction. The fourth series is an initial-state control in which the same model starts from $|g\rangle$ instead of $|e\rangle$, with the reference frame matched to $\chi_g$. (b)~Bath-model dependence of $p_2$, $p_3$, and $p_4$, shown as ratios to the baseline bath. Adjacent-only $T_1$ uses the same weights but permits only $k\to k-1$ transitions and reproduces the baseline to all displayed digits (ratio $1.0000$). Ohmic $T_1$ weights the rates by transition frequency, $\Gamma_{k\to j}\propto \omega_{kj}|\langle j|\hat{n}|k\rangle|^{2}$, and frequency-weighted dephasing replaces the linear diagonal by $d_k=(E_k-E_0)/(E_1-E_0)$. (c)~Fraction of $p_2$ (dark bars) and $p_3$ (light bars) that remains when the dominant channel is removed separately within each bath model. All comparisons use the RWA open-system model.}
    \label{fig:channels}
\end{figure*}

We next ask which dissipative process produces the persistent population within the same model. During the drive, reversible dressing transfers population into higher levels that would return adiabatically if the evolution remained coherent. Decoherence converts part of this dressed population into irreversible bare-level occupation. To identify the relevant channel, Fig.~\ref{fig:channels}(a) removes one dissipator at a time, with each case propagated independently rather than obtained by subtraction. Without pure dephasing, $p_{2}$ falls to $0.251$ of its full-model value while $p_{3}$ remains at $0.963$. Without relaxation, $p_{2}$ remains at $0.770$, but $p_{3}$ falls to $0.034$ and $p_{4}$ to $1.4\times10^{-3}$. Thus $p_{2}$ is produced mainly by dephasing, whereas $p_{3}$ and higher levels are produced mainly by relaxation. Removing both channels nearly eliminates the ladder, leaving $0.015$ of $p_{2}$ and $1.6\times10^{-4}$ of $p_{3}$.

The fourth series in Fig.~\ref{fig:channels}(a) is an initial-state control. We run the same full model from $\ket{g}$ rather than $\ket{e}$ and match the reference frame to the ground-state pull $\chi_{g}$. The remaining populations are approximately $0.011$ of $p_{2}$ and $0.023$ of $p_{3}$ relative to the excited-state run. Because the $\chi_{g}$-matched calculation was not independently refined along a separate Fock ladder, these ratios are indicative rather than converged. The control nevertheless shows that the observed population is associated with the excited-state manifold rather than generic drive heating from any initial state.

Because the channel weights depend on the bath model, we also test whether the channel assignment changes when the dissipators are modified. Figure~\ref{fig:channels}(b) compares several bath choices. An adjacent-only $\Tone$ reproduces the baseline to all displayed digits because the flat bath is already effectively adjacent-only. An Ohmic $\Tone$, which weights each rate by its transition frequency, changes $p_{2}$, $p_{3}$, and $p_{4}$ to $0.983$, $0.893$, and $0.845$ of their baseline values. Frequency-weighted dephasing changes them to $0.905$, $0.976$, and $0.997$. The absolute populations therefore vary by up to about $15\%$, but the channel assignment does not. Figure~\ref{fig:channels}(c) repeats the dominant channel removal within each bath model. Removing $\Tone$ leaves $0.034$ of $p_{3}$ in the baseline bath and $0.038$ in the Ohmic bath, while removing $\Tphi$ leaves $0.251$ of $p_{2}$ in the baseline bath and $0.277$ with frequency-weighted dephasing. Across the RWA open-system models tested here, $p_{2}$ remains associated primarily with dephasing and $p_{3}$ with relaxation. The controls do not identify the origin of the small residue that remains when both channels are removed.

\FloatBarrier
\section{Limits of the excitation-conserving approximation}
\label{sec:limits}

\begin{figure*}[t]
    \centering
    \includegraphics[width=\textwidth]{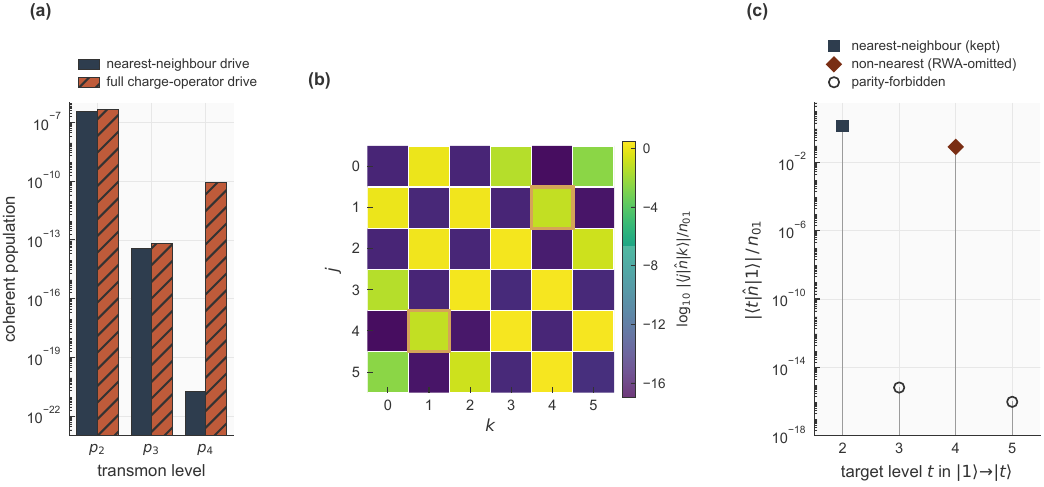}
    \caption{Coherent test of the excitation-conserving approximation and the charge-operator terms that it omits. (a)~Populations for $k=2$ to $4$ in a dissipation-free transmon-only benchmark, comparing the nearest-neighbour RWA drive (solid dark bars) with the full charge-operator drive including the counter-rotating classical-drive terms (hatched light bars), at $\bar{n}=2$ ($\bar{n}/n_{\mathrm{crit}}=0.026$) and $t=12$~ns. \emph{The transmon propagation is unitary and contains no Lindblad dissipators; the cavity field enters only through the prescribed classical trajectory. This panel tests the coherent drive truncation and is not a beyond-RWA open-system population prediction.} (b)~Magnitude of the charge-operator matrix elements at $n_g=0$, $\log_{10}\left(|\langle j|\hat{n}|k\rangle|/n_{01}\right)$ with $n_{01}=|\langle 0|\hat{n}|1\rangle|$, for $j,k=0$ to $5$. Nonzero elements occur only for odd $j+k$; even-$(j+k)$ elements are below $10^{-12}$ (darkest cells). White outlines mark parity-allowed elements and gold outlines mark the non-nearest-neighbour pair $\langle 4|\hat{n}|1\rangle$, $\langle 1|\hat{n}|4\rangle$. (c)~First-order coupling from $|1\rangle$, $|\langle t|\hat{n}|1\rangle|/n_{01}$ for $t=2$ to $5$, on a logarithmic axis. The filled square marks the nearest-neighbour parity-allowed element retained by the RWA drive, the filled diamond marks the parity-allowed non-nearest-neighbour element $\langle 4|\hat{n}|1\rangle$ omitted by the $\Delta k=+1$ drive operator, and open circles are parity-forbidden. The RWA drive reaches $|4\rangle$ only through three successive rungs. Panels~(b) and~(c) establish the omitted matrix element but do not divide the panel-(a) discrepancy among non-nearest-neighbour elements, counter-rotating coupling, and counter-rotating drive.}
    \label{fig:coherent_rwa}
\end{figure*}

The results in Secs.~\ref{sec:dynamics} and \ref{sec:mechanism} belong to the excitation-conserving model. To test one coherent source of approximation error, we compare the nearest-neighbour rotating-drive truncation with a full charge-operator drive in a transmon-only unitary calculation using a prescribed classical cavity trajectory. No dissipator acts on the transmon in this benchmark, so any difference comes from the coherent drive truncation rather than from the open-system construction. Figure~\ref{fig:coherent_rwa}(a) shows the low-occupation test at $\bar{n}=2$ ($\bar{n}/\ncrit = 0.026$) and $t=12$~ns, a regime where the RWA would normally be expected to be very accurate. In this regime, the beyond-RWA propagator agrees with a dense lab-frame unitary reference to $1.74\times10^{-11}$.

The nearest-neighbour drive already underestimates the full charge-operator populations in this low-occupation test, and the discrepancy grows with level index. The ratios $p_{k}^{\mathrm{RWA}}/p_{k}^{\mathrm{full}}$ are $0.74$ for $k=2$, $0.54$ for $k=3$, and $2.1\times10^{-11}$ for $k=4$. We then repeat the coherent benchmark at the strong-drive point, $\nlp=80$, using the same $\chi_e$-conditioned damped semiclassical cavity trajectory that defines the readout drive. The comparison is made at drive switch-off, $t=260$~ns, while the coherent higher-level populations are still appreciable. With $n_{\mathrm{lev}}=12$ and $\mathrm{d}t=0.002$~ns, the full charge-operator calculation gives $p_2=0.261295$ and $p_3=0.034053$, whereas the nearest-neighbour RWA drive gives $p_2=0.236527$ and $p_3=0.025830$. Thus
\begin{equation}
    \frac{p_2^{\mathrm{RWA}}}{p_2^{\mathrm{full}}}=0.905,
    \qquad
    \frac{p_3^{\mathrm{RWA}}}{p_3^{\mathrm{full}}}=0.759.
\end{equation}
The strong-drive comparison is numerically stable. Increasing $n_{\mathrm{lev}}$ from $10$ to $12$ changes the full-model $p_2$ and $p_3$ by less than $0.006\%$, and reducing the timestep from $0.004$ to $0.002$~ns changes them by approximately $0.01\%$. The norm of the full-model propagation is conserved to $5.2\times10^{-6}$ over the pulse. The discrepancy grows further at deeper levels: at the same time, $p_4^{\mathrm{RWA}}/p_4^{\mathrm{full}}=0.544$ and $p_5^{\mathrm{RWA}}/p_5^{\mathrm{full}}=0.0347$. These ratios describe only the coherent drive-Hamiltonian truncation. We do not use them as correction factors for the dissipative open-system populations, and they do not bound the error on those populations. At $t_f=619.3$~ns, the coherent residues are $p_2=2.95\times10^{-6}$ and $p_3=7.78\times10^{-11}$, far below the reported open-system values. This is consistent with the channel-removal result that the reported residual populations are generated mainly by dissipation during the driven evolution rather than by coherent population that survives to the endpoint. None of the calculations presented here determines the RWA error on the dissipative populations themselves.

Panels (b) and (c) show one structural difference that becomes important for $k=4$. At $n_g=0$, the transmon charge operator obeys a parity rule: $\melem{j}{\nhat}{k}$ is nonzero only when $j+k$ is odd, while all even-$(j+k)$ elements are below $10^{-12}$. For first-order coupling out of $\ket{1}$, the nearest-neighbour element $\melem{2}{\nhat}{1}/n_{01}=1.370$ is large. The $t=3$ and $t=5$ elements are parity-forbidden and negligible, at $6.6\times10^{-16}$ and $9.6\times10^{-17}$, whereas $\melem{4}{\nhat}{1}/n_{01}=0.0796$ is smaller but parity-allowed and non-nearest-neighbour. Because the excitation-conserving drive operator retains only $\Delta k = +1$ terms, it removes the direct $\ket{1}\!\to\!\ket{4}$ coupling and can reach $\ket{4}$ only through three successive nearest-neighbour steps. This omitted direct path is one structural difference relevant to the strong suppression of the RWA $p_{4}$ population in Fig.~\ref{fig:coherent_rwa}(a).

Panels (b) and (c) establish that the $\melem{4}{\nhat}{1}$ element is omitted by the nearest-neighbour drive. They do not determine how much of the discrepancy in panel (a) comes from non-nearest-neighbour matrix elements, counter-rotating coupling, or counter-rotating drive. That separation would require an ablation calculation that was not performed here.

\FloatBarrier
\section{Beyond-RWA open-system description and its failure}
\label{sec:failure}

\begin{figure*}[t]
    \centering
    \includegraphics[width=\textwidth]{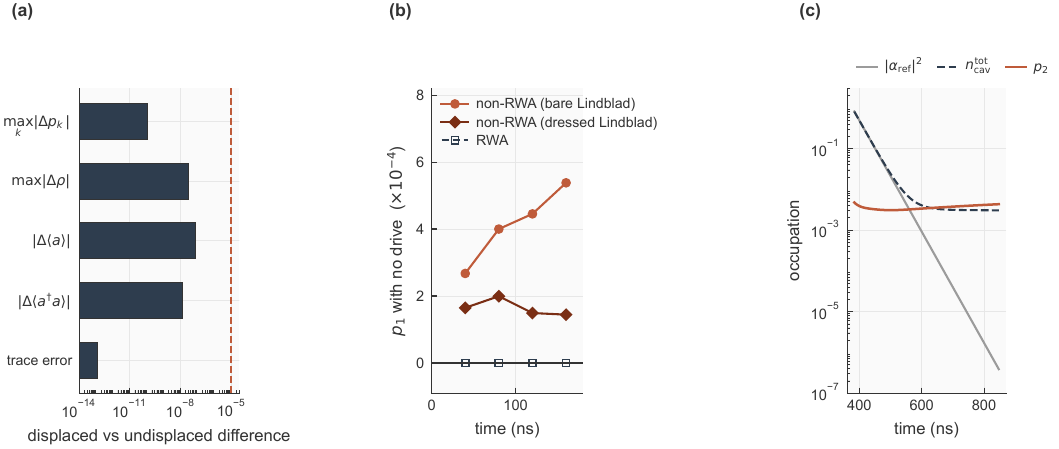}
    \caption{Consistency tests for the beyond-RWA open-system model. (a)~Difference between two implementations of the same model: a displaced-frame calculation with $N_f=10$ and an undisplaced calculation in the same cavity-rotating frame with $N_f=18$. Both use $5$ transmon levels, $\bar{n}=4$, $\mathrm{d}t=0.004$~ns, and all dissipators, and are compared at $t=24$~ns. From top: $\max_k|\Delta p_k|$, $\max|\Delta\rho|$, $|\Delta\langle a\rangle|$, $|\Delta\langle a^\dagger a\rangle|$, and the trace deviation from unity. The dashed line marks the $10^{-5}$ acceptance tolerance. Agreement confirms implementation consistency but not physical validity. (b)~Zero-drive ground-state test. The system starts from $\lvert g,0\rangle$ with $\varepsilon=0$ at all times ($6$ transmon levels, $N_f=8$, charge cutoff $25$, $\mathrm{d}t=0.008$~ns). For the beyond-RWA model this is a diagnostic initial condition because the true invariant state is the dressed ground state, not the bare product state. Open squares show the RWA model, for which $p_1$ remains identically zero. Filled circles show the beyond-RWA model with bare-basis Lindblad dissipators, for which $p_1$ grows. Filled diamonds show the same model with dressed-basis (secular) dissipators, for which $p_1$ stays bounded and does not grow. (c)~Beyond-RWA open-system run at $\bar{n}_{\mathrm{LP}}=40$ followed by a $250$~ns wait: reference cavity occupation $|\alpha_{\mathrm{ref}}|^{2}$ (solid grey), total cavity occupation $n_{\mathrm{cav}}^{\mathrm{tot}}$ (dashed dark), and $p_2$ (solid light). Long after switch-off, $|\alpha_{\mathrm{ref}}|^{2}$ decays while $n_{\mathrm{cav}}^{\mathrm{tot}}$ approaches a floor and $p_2$ continues to rise.}
    \label{fig:vacuum_gate}
\end{figure*}

Section~\ref{sec:limits} shows that the excitation-conserving Hamiltonian is not quantitatively reliable for the higher levels. The natural extension is to retain the counter-rotating terms and propagate the open-system dynamics with the full Hamiltonian. In its naive form, however, that extension fails a physical-consistency test. Figure~\ref{fig:vacuum_gate} shows the failure and why it can be missed by ordinary numerical validation.

Panel (a) shows a numerical consistency check that passes. Two independent implementations of the \emph{same} beyond-RWA open-system model, one displaced with $N_{f}=10$ and one undisplaced in the same cavity-rotating frame with $N_{f}=18$, agree with all dissipators active. The differences are $\max_{k}\lvert\Delta \pk\rvert = 1.03\times10^{-10}$ in the populations, $2.87\times10^{-8}$ in the density-matrix elements, $7.42\times10^{-8}$ in the cavity amplitude, and $1.34\times10^{-8}$ in the cavity occupation, with a trace deviation of $1.23\times10^{-13}$ against a $10^{-5}$ tolerance. This confirms that the two implementations solve the same model consistently. It does not establish that the model is physically valid because both use the same dissipator construction and therefore share the same defect.

Panel (b) exposes that defect. At zero temperature and with the drive off, a physically consistent model should not sustain excitation away from the ground state of the interacting Hamiltonian. For the excitation-conserving model, this ground state is exactly $\ket{g,0}=\ket{g}\otimes\ket{0}$. Starting from this state with $\varepsilon = 0$, the model keeps $p_{1}$ identically zero at every sampled time. The beyond-RWA case is different because its true ground state is dressed rather than the bare product state $\ket{g,0}$. We therefore use $\ket{g,0}$ only as a diagnostic initial condition. With bare-basis Lindblad dissipators, $p_{1}$ grows from $2.67\times10^{-4}$ at $40$~ns to $5.38\times10^{-4}$ at $160$~ns, corresponding to roughly $2.3\times10^{-6}$ per ns after the transient. This sustained secular pumping is not the bounded response expected from the initial mismatch between $\ket{g,0}$ and the dressed ground state. The bare-basis jump operators violate detailed balance about the dressed ground state and continuously pump the system. The effect is therefore a defect of the dissipator construction, not readout-induced excitation or a coding error. Stated in the physically relevant basis, the criterion is simple: when initialized in the dressed ground state, a consistent zero-drive model must leave that state stationary.

Panel (c) shows the consequence in a driven calculation. A beyond-RWA open-system run at $\nlp = 40$, followed by a $250$~ns wait, should approach its steady state after the drive has been off for a long time. Instead, the reference occupation $\lvert\aref\rvert^{2}$ decays to $3.9\times10^{-7}$ while the total cavity occupation approaches a floor of $3.05\times10^{-3}$ and $p_{2}$ continues to grow. Across the wait, $p_{2}$ increases by $28.8\%$ and $p_{3}$ by $28.3\%$. The cavity floor is not itself the problem; it is the physical dispersive photon cloud retained by the displaced frame. The failure is the continued increase of $p_{2}$ after the drive is long off, which is the spurious pumping seen in panel (b) accumulating into an unphysical late-time rise. We therefore do not interpret any population from the beyond-RWA open-system runs as physical. They are used only as diagnostics, and the excluded strong-drive case is shown separately in Fig.~\ref{fig:supp_nonrwa_diagnostic}.

The agreement between the displaced and undisplaced calculations in panel (a) would normally be a strong numerical validation, and here it reaches the $10^{-10}$ level. It is nevertheless blind to this problem because it compares two implementations of the same physical model. The zero-drive ground-state test in panel (b) provides the additional check needed to distinguish a consistent model from an inconsistent one. We therefore use it as a validity gate for bare-basis open-system treatments that retain counter-rotating terms.

The same analysis also indicates how to remove the defect. We express the relaxation and cavity jump operators in the eigenbasis of the full counter-rotating Hamiltonian and retain only strictly downward transitions between eigenstates. This dressed-basis, or secular, construction makes the dressed ground state a dark state of every dissipator. We test it in Fig.~\ref{fig:vacuum_gate}(b). Starting both models from $\lvert g,0\rangle$ as a diagnostic initial state, the bare-basis model again shows spurious growth, with $p_{1}$ increasing from $2.67\times10^{-4}$ to $5.38\times10^{-4}$ at a rate of $2.3\times10^{-6}$ per ns. In the dressed-basis model, $p_{1}$ remains bounded, with a slope of $-1.7\times10^{-7}$ per ns. More importantly, when the dressed model is initialized in the dressed ground state, that state remains stationary to below $10^{-11}$ over $160$~ns. The failure therefore comes from the bare-basis dissipators rather than the counter-rotating Hamiltonian, and the secular construction removes it at zero drive. We have not used these dissipators to regenerate the driven production populations reported above. The construction implemented here uses the static dressed basis, which is appropriate for the zero-drive test. In the driven beyond-RWA problem, the counter-rotating and non-nearest-neighbour terms remain explicitly time dependent in the rotating-frame representations considered here. A dissipative treatment that remains consistent under the strong periodic drive therefore requires a Floquet or comparably time-dependent dressed construction. The difficult part is not propagating a time-dependent Hamiltonian, but constructing dissipation that is consistent with the driven dressed states.

\FloatBarrier
\section{Discussion}\label{sec:discussion}

All conclusions that remain after these tests are confined to the excitation-conserving open-system model. Within that model, dispersive readout at $\nlp = 80$ leaves persistent population in the first few transmon levels, at about $10^{-3}$ for $p_{2}$ and $10^{-5}$ for $p_{3}$. The post-readout decay follows the level-relaxation rates rather than the cavity ring-down, identifying the population as bare-level occupation within the model. The channel-removal calculations further show a level-dependent dissipative origin: $p_{2}$ is produced mainly by dephasing, whereas $p_{3}$ and higher levels are produced mainly by relaxation. This assignment remains the same for the bath variants tested. These statements are converged predictions of the model, not exact results for the full driven open system.

We do not yet have the corresponding population values from a validated beyond-RWA open-system model, and the coherent benchmarks in Sec.~\ref{sec:limits} show why this limitation matters. The nearest-neighbour drive truncation omits parity-allowed non-nearest-neighbour charge couplings and becomes increasingly inaccurate for deeper levels. Simply restoring the full Hamiltonian while retaining bare-basis Lindblad dissipators does not solve the problem because that model fails the zero-drive consistency test in Sec.~\ref{sec:failure}. We therefore treat those beyond-RWA open-system calculations only as a diagnostic negative result: the bare-basis dissipator construction is ruled out, its failure is identified, and a dressed-basis repair is implemented. The dressed-basis model passes the zero-drive ground-state test in Sec.~\ref{sec:failure} when initialized in the dressed ground state. Extending this construction to regenerate the driven populations requires a dissipator that is consistent with the time-dependent dressed states, which we leave for future work. A direct lab-frame calculation with static dressed dissipators is practical at low occupation but becomes computationally prohibitive at the readout occupations considered here with the present numerical architecture because the cavity basis must resolve the full coherent field. We therefore do not use that calculation to infer strong-drive populations.

Two limitations set the quantitative range of the converged RWA results. First, the strong-drive case has $\nlp/\ncrit \simeq 1.03$, so it lies near the critical photon number rather than deep in the dispersive regime. The coherent comparisons in Sec.~\ref{sec:limits} test the nearest-neighbour drive truncation at both low occupation and the $\nlp=80$ drive strength. At the strong-drive point, the discrepancies are substantial and level dependent, but they do not determine the RWA error on the reported dissipative populations and cannot be used as a correction factor. Second, the remaining uncertainty within the RWA model is dominated by the bath choice rather than by numerical convergence. The bath-model spread in $p_{2}$, $p_{3}$, and $p_{4}$ reaches $9.5$, $10.7$, and $15.5\%$, respectively, which is larger than the sub-percent-to-percent numerical uncertainty shown in Fig.~\ref{fig:supp_convergence}(b). Despite this spread in the absolute populations, the channel assignment remains unchanged.

\FloatBarrier
\section{Conclusion}\label{sec:conclusion}

We have calculated the level-resolved transmon population produced by dispersive readout in a converged excitation-conserving open-system model. Within this model, part of the population remains after the drive as bare-level occupation. Channel-removal tests trace this residual population to decoherence acting on the drive-dressed transmon, with a level-dependent assignment that is stable across the bath models considered. We also test where the model ceases to be quantitatively reliable. Dissipation-free transmon-only benchmarks at low occupation and at the $\nlp=80$ drive strength show the error associated with the nearest-neighbour coherent-drive truncation and identify a parity-allowed charge matrix element omitted by that operator. A separate zero-drive ground-state test rules out the naive beyond-RWA open-system extension with bare-basis dissipators. Replacing them with dressed-basis dissipators removes this zero-drive inconsistency and leaves the dressed ground state stationary. The result is therefore a converged but scope-limited prediction, together with explicit tests of the approximation on which that prediction depends.

\FloatBarrier
\appendix

\section{Numerical convergence and uncertainty budget}
\label{app:convergence}

\begin{figure*}[t]
    \centering
    \includegraphics[width=\textwidth]{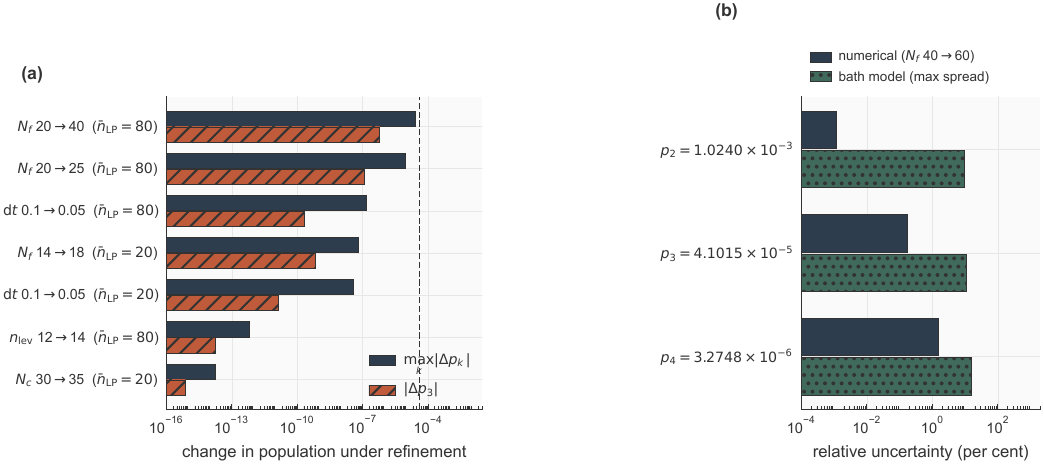}
    \caption{Numerical convergence and uncertainty budget for the RWA results. (a)~Change in the final populations under each refinement, https://www.overleaf.com/projectshown on a logarithmic axis. Dark bars give $\max_k|\Delta p_k|$ over all levels and hatched light bars give $|\Delta p_3|$. The labels specify the refinement and run: Fock cutoff $N_f$ ($20\to25$ and $20\to40$ at $\bar{n}_{\mathrm{LP}}=80$; $14\to18$ at $\bar{n}_{\mathrm{LP}}=20$), timestep $\mathrm{d}t$ ($0.1\to0.05$~ns at both powers), transmon levels $n_{\mathrm{lev}}$ ($12\to14$), and charge-basis cutoff $N_c$ ($30\to35$). The coarser calculation is the production baseline in each comparison. The dashed vertical line shows the production value $p_3 = 4.04\times10^{-5}$ at $\bar{n}_{\mathrm{LP}}=80$ for scale; the converged $N_f=60$ value is $4.1015\times10^{-5}$. (b)~Relative uncertainty in $p_2$, $p_3$, and $p_4$ at $\bar{n}_{\mathrm{LP}}=80$, expressed as a percentage of the central value and shown on a logarithmic axis; the tick labels give the converged $N_f=60$ populations. Dark bars show numerical uncertainty from the $N_f\,40\to60$ step, while hatched light bars show bath-model sensitivity, defined as the largest relative deviation among the three alternative baths in Fig.~\ref{fig:channels}(b). Both panels refer only to the RWA model and neither gives a beyond-RWA correction.}
    \label{fig:supp_convergence}
\end{figure*}

Figure~\ref{fig:supp_convergence} summarizes the convergence data discussed in Sec.~\ref{sec:scope}. Panel (a) gives the change in the final populations for each of the seven refinement comparisons. The largest change, $2.68\times10^{-5}$ from the $N_{f}\,20\to40$ refinement, defines the conservative global resolution threshold used in Fig.~\ref{fig:final_populations}(a). For $p_{3}$, every refinement changes the population by at most $5.9\times10^{-7}$. Panel (b) separates numerical uncertainty from bath-model sensitivity for $p_{2}$, $p_{3}$, and $p_{4}$. Across the $N_{f}\,40\to60$ step, the numerical contributions are $0.0012$, $0.169$, and $1.53\%$, whereas the bath-model spreads are $9.48$, $10.73$, and $15.46\%$. At each level, the bath-model sensitivity is larger than the residual numerical uncertainty.

\section{Beyond-RWA open-system diagnostics}
\label{app:nonrwa}

\begin{figure*}[t]
    \centering
    \includegraphics[width=\textwidth]{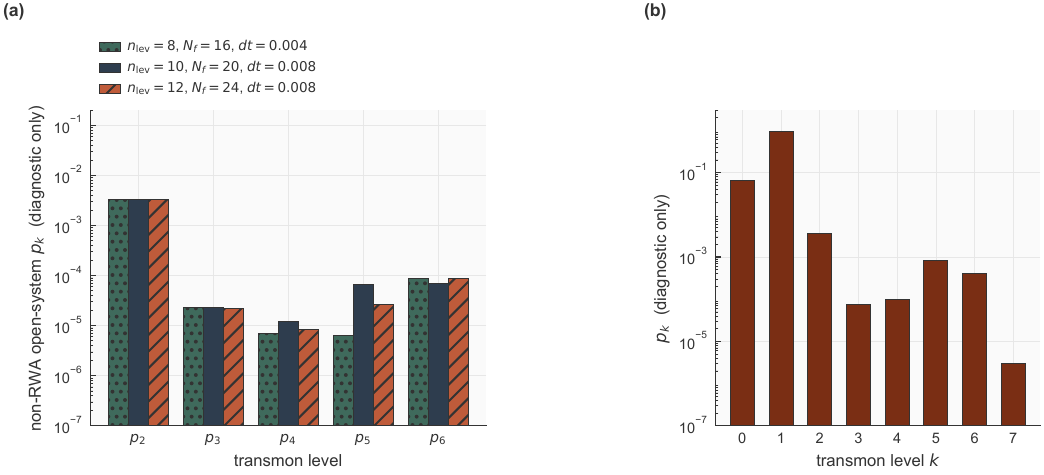}
    \caption{Beyond-RWA open-system calculations used only as model-validity diagnostics. \emph{No physical level population, ratio, or correction factor is inferred from either panel}; these calculations use bare-basis Lindblad dissipators and fail the zero-drive ground-state consistency test in Fig.~\ref{fig:vacuum_gate}(b). (a)~Truncation ladder at $\bar{n}_{\mathrm{LP}}=40$: final populations $p_2$ to $p_6$ for $n_{\mathrm{lev}}=8$, $N_f=16$, $\mathrm{d}t=0.004$~ns; $n_{\mathrm{lev}}=10$, $N_f=20$, $\mathrm{d}t=0.008$~ns; and $n_{\mathrm{lev}}=12$, $N_f=24$, $\mathrm{d}t=0.008$~ns. The timestep is not fixed across the ladder and is listed for each case. Under successive refinement, $p_2$ changes by $+0.33$ and $-0.58$~per~cent and $p_3$ by $-0.62$ and $-2.51$~per~cent, within the $5$~per~cent tolerance. In contrast, $p_4$ changes by $+77.8$ and $-31.4$~per~cent, $p_5$ by $+981$ and $-61.5$~per~cent, and the combined bin $P(k\ge4)$ by $+49.5$ and $-18.1$~per~cent, all outside the $10$~per~cent tolerance. (b)~Excluded strong-drive calculation at $\bar{n}_{\mathrm{LP}}=80$ with $n_{\mathrm{lev}}=8$, $N_f=20$, and $\mathrm{d}t=0.004$~ns. Population accumulates near the top of the retained ladder: $p_5=8.5\times10^{-4}$ and $p_6=4.2\times10^{-4}$ exceed $p_3=7.4\times10^{-5}$ and $p_4=1.0\times10^{-4}$ by factors of roughly $4$ to $11$, reversing the ordering expected for a converged ladder.}
    \label{fig:supp_nonrwa_diagnostic}
\end{figure*}

Figure~\ref{fig:supp_nonrwa_diagnostic} gives two additional reasons, separate from the ground-state consistency failure in Sec.~\ref{sec:failure}, that the beyond-RWA open-system runs cannot be interpreted as physical populations. In panel (a), the truncation ladder at $\nlp = 40$ converges for $p_{2}$ and $p_{3}$ within the $5\%$ tolerance, but not for $p_{4}$ and higher levels, which change by tens to hundreds of percent. The combined bin $P(k\ge4)$ also fails the criterion, changing by $18.1\%$, so coarse graining does not provide a converged alternative. Panel (b) shows the excluded strong-drive case. With only eight transmon levels retained at $\nlp = 80$, population accumulates near the top of the truncated ladder: $p_{5}=8.5\times10^{-4}$ and $p_{6}=4.2\times10^{-4}$ exceed $p_{3}$ and $p_{4}$ by about an order of magnitude and reverse the expected ladder ordering. None of these values is used in any physical claim in this paper.

\FloatBarrier

\bibliography{references}

\end{document}